\documentclass[10pt]{article}
\usepackage{hyperref}
\usepackage{color,graphicx}
\begin{document}

\begin{center}

\textbf{\LARGE On arc-disjoint Hamiltonian cycles in De Bruijn graphs}

\bigskip{\large \href{http://www.ms.sapientia.ro/~kasa}{Z. K\'asa}}

\href{http://www.emte.ro}{Sapientia Hungarian University of Transylvania}

email: \texttt{kasa@ms.sapientia.ro}

\bigskip \emph{Presented at:} \\
\href{http://riesz.math.klte.hu/~macs/abstracts.pdf}{5th Joint Conference on Mathematics and Computer Science, Debrecen, June 9-12, 2004} \\
 \emph{and  published in Hungarian:}\\ De Bruijn-gr\'afok mint h\'al\'azati modellek (De Bruijn Graphs as Network Model), \href{http://www.emt.ro/downloads/muszaki_szemle/msz43.pdf}{M\H{u}szaki Szemle/Technical Review, 43 (2008) 3-6} 
\end{center}

\bigskip
{\small
\noindent\textbf{Abstract.} We give two equivalent formulations of a conjecture [2,4] on the number of arc-disjoint Hamiltonian cycles in De Bruijn graphs. 
}

\bigskip\noindent
A De Bruijn word  of type $(q,k)$ for a given $q$ and $k$ is a word over an alphabet with $q$ letters, containing all  $k$-length words exactly once. The length of such a word is $q^k+k-1$. 
For example if $q=3, k=2$, then  0012202110 is a De Bruijn word of type $(3,2)$.

\medskip\noindent
For a $q$-letter alphabet ${\cal A}$ the De Bruijn graph $B(q,k)$ is defined as:\\
${B(q,k)= (V(q,k), E(q,k))}$
with \\
$\bullet$ {$V(q,k)={\cal A}^k$}   \quad the set of vertices
\\
$\bullet$ {$E(q,k)= {\cal A}^{k+1}$}  the set of directed arcs
\\
$\bullet$ there is an arc from vertex {$x_1x_2\ldots x_k$} to  vertex {$y_1y_2\ldots y_k$}  if 
{$x_2x_3\ldots x_k=y_1y_2\ldots y_{k-1}$} 
and this arc is denoted by {$x_1x_2\ldots x_ky_k$}. 
      
\medskip\noindent
In the De Bruijn graph $B(q,k)$ a path (i. e. a walk with distinct vertices)
 $ {a_1a_2\ldots a_k,}$ \  $  {a_2a_3\ldots a_{k+1},}$ \  
\ldots \ ${ a_{r-k+1}a_{r-k+2}\ldots a_r , \quad r>k}$ \   
corresponds to an $r$-length word 
 {$a_1a_2\ldots a_ka_{k+1}\ldots a_r$},
 which 
is obtained by maximal overlapping of the neighboring vertices. 

\begin{figure}[t]
\centering\includegraphics[scale=0.6]{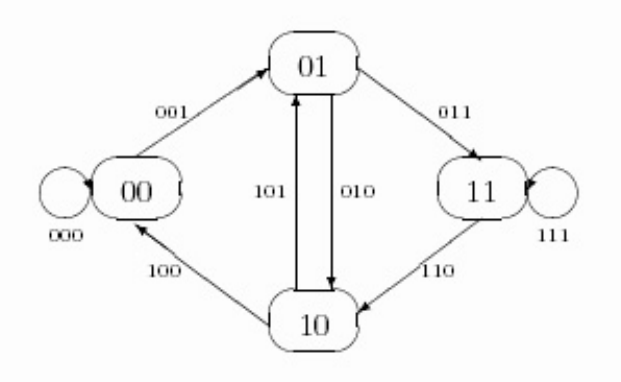}
\label{br22}\caption{De Bruijn graph $B(2,2)$} 
\end{figure}

\medskip\noindent
In $B(2,3)$ the path {$001$, $010$, $101$} corresponds to the 
word {$00101$.}  

\medskip\noindent
Every maximal length path in the graph $B(q,k)$ (which is a
Hamiltonian one) corresponds to a De Bruijn word.

\bigskip\noindent{\textcolor{red}{Algorithm to generate all De Bruijn words}}
\\
$a_1,a_2, \ldots, a_q$ the letters of an alphabet, with the values: $a_i=i-1$ for $i=1,2,\ldots ,q$, 
$S$ a vector to store the letters of a De Bruijn word, 
$B$ a vector to store the states of words.  $val(S_{i-k}\ldots  S_{i-1})$ represents the value of the number $S_{i-k}\ldots S_{i-1}$ in base  $q$. 

\begin{figure}[t]
\centering\includegraphics[scale=0.6]{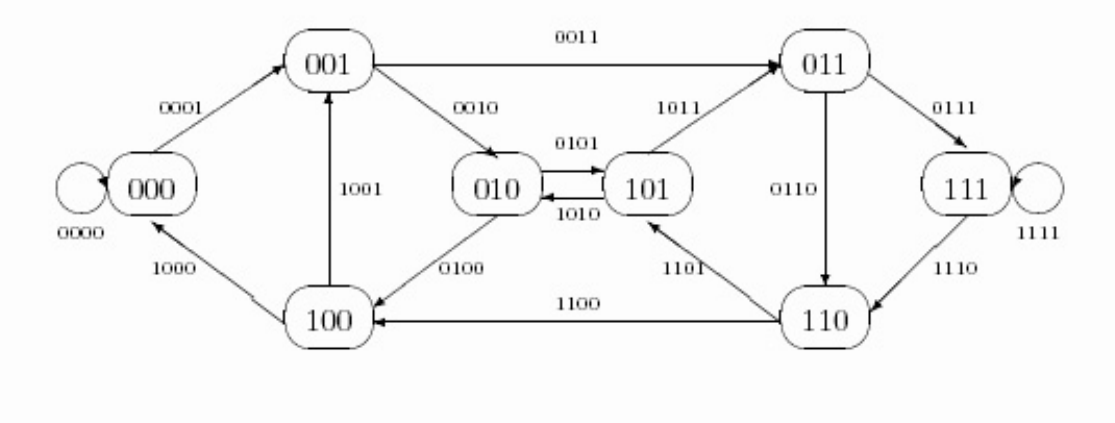}
\label{br23}\caption{De Bruijn graph $B(2,3)$} 
\end{figure}

\medskip\noindent Initially: $S_i :=0$ for all $i=1,2,\ldots, k$, 
$B_0:=1$ and $B_i:=0$ for other indices. 
The call of the procedure: DeBruijn (S,k+1,k,B).

\begin{tabbing}
\textbf{{procedure}} DeBruijn ($S,i,k,B$)\\
\textbf{{for}} \= $j:=1$ \textbf{{to}} $q$ \textbf{{do}}\\
             \> $S_i := a_j$\\
             \>  $r := val(S_{i-k+1}S_{i-k+2} \ldots S_{i}) $\\
             \> \textbf{{if}} $B_r=0$ \=\textbf{{then}} \= $B_r:=1$\\
             \>                     \>              \> DeBruijn($S,i+1,k,B$)\\
             \>                     \>              \> $B_r:=0$\\
             \>                     \>\textbf{{else}} \= \textbf{{if}}  \=         $length(S)=q^k+k-1$\\ 
            \> \> \> \> \textbf{{then}} \= write (S)\\
             \> \> \> \>                                           \>  exit for\\
             \>         \>            \>               \> \textbf{{endif}}  \\          
             \> \textbf{{endif}}\\
\textbf{{endfor}} \\
\textbf{{endprocedure}}            
\end{tabbing}

\begin{figure}[t]
\centering\includegraphics[scale=0.8]{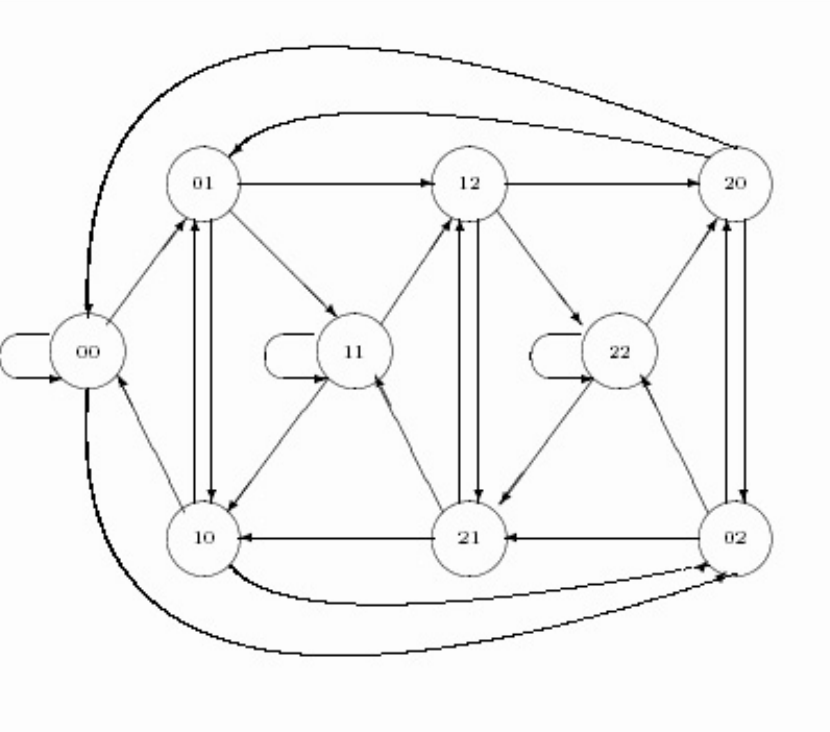}
\label{br32}\caption{De Bruijn graph $B(3,2)$} 
\end{figure}

\medskip\noindent 
In the directed graph $B(q,k)$ there always exists an Eulerian circuit because
it is connected and  all its
vertices have the same indegree and outdegree $q$. An Eulerian
circuit in $B(q,k)$  is a Hamiltonian path in $B(q,k+1)$ (which always can
be continued in a Hamiltonian cycle).

\medskip\noindent\textbf{{Conjecture 1.}} [2, 4]

\noindent\textcolor{red}{In the De Bruijn graph $B(q,k)$ for $q\ge 2$ and $k>1$ the number of arc-disjoint Hamiltonian cycles is $q-1$.}

\medskip\noindent
Let us define a morphism $\mu$ on words over an alphabet $A=\{0,1,\ldots, q-1\}$:
\begin{eqnarray*}
\mu(0)   &=& 0\\
\mu(i)   &=& i+1, \quad \textrm{ if } 1\le i <q-1\\
\mu(q-1) &=& 1 \\
\end{eqnarray*}
It is easy to see that $\mu^{q-1}(u)=u$ for any $u\in A^*$.

\medskip\noindent\textbf{Example:} 

Let be a word $w=00102113230331220$ on the alphabet $A=\{0,1,2,3\}$.

$\mu(w)\;\, ={00203221310112330}$ 

$\mu^2(w)={00301332120223110}$

\medskip\noindent
From a De Bruijn word we obtian a Hamiltonian cycle. Let an $H_0$ be such a Hamiltonian cycle. Using $\mu^k$ for $k=1,2,\ldots q-2$ we will obtain De Bruijn words corresponding to the Hamiltonian cycles $H_1, H_2, \ldots, H_{q-2}$.

\medskip\noindent\textbf{{Conjecture 2.}}

\noindent\textcolor{red}{In the De Bruijn graph $B(q,k)$ for {$q>2, k>1$} there exists a Hamiltonian cycle {$H_0$} such that the Hamiltonian cycles  $H_1$, $H_2$, \ldots $H_{q-2}$ (obtained from {$H_0$} by using the morphisms $\mu_k$, $k=1,2,\ldots q-2$)), together with $H_0$ are  arc-disjoint Hamiltonian cycles.}

\medskip\noindent 
{$B(3,2)$}\\
$H_0:$ {0011220210}, \\ $H_1:$  {0022110120}\\
{$B(3,3)$} \\
$H_0:$ {00010021011022202012111221200}, \\
$H_1:$ {00020012022011101021222112100}
\\
{$B(4,2)$}\\  $H_0:$ {00102113230331220}, \\ 
$H_1:$ {00203221310112330}, \\
$H_2:$ {00301332120223110}  
\\
{$B(5,2)$} \\
 $H_0:$ {00102112041422430332313440},\\  
$H_1:$ {00203223012133140443424110},  \\
 $H_2:$ {00304334023244210114131220}, \\
 $H_3:$ {00401441034311320221242330}

\medskip\noindent
Two words $u$ and $v$ are equivalent if $v=\mu^k(u)$ for some $1\le k\le q-2$ 

\medskip\noindent 
If $v=\mu^k(u)$, $1\le k\le q-2$, then $u=\mu^{q-1-k}(v)$ because, applying $\mu^{q-1-k}$ to $v=\mu^k(u)$ we obtain
$\mu^{q-1-k}(v)=\mu^{q-1-k}\big(\mu^k(u)\big) = \mu^{q-1}(u)=u.$

\medskip\noindent
In the set of all De Bruijn words over an alphabet $A=\{0,1,\ldots, q-1\}$ this equivalence relation will introduce a partition of De Bruijn words in equivalent classes.

\medskip\noindent
\textbf{{An equivalent assertion to Conjecture 2.}}

\medskip\noindent\textcolor{red}{In the De Bruijn graph {$B(q,k)$} for {$q>2, k>1$} there exist the Hamiltonian cycles  $H_0$, $H_1$, $H_2$, \ldots $H_{q-2}$ such that they correspond to De Bruijn words $B_0$, $B_1$, $B_2$, \ldots $B_{q-2}$ from the same equivalence class.}

\begin{figure}
\centering\includegraphics[scale=0.5]{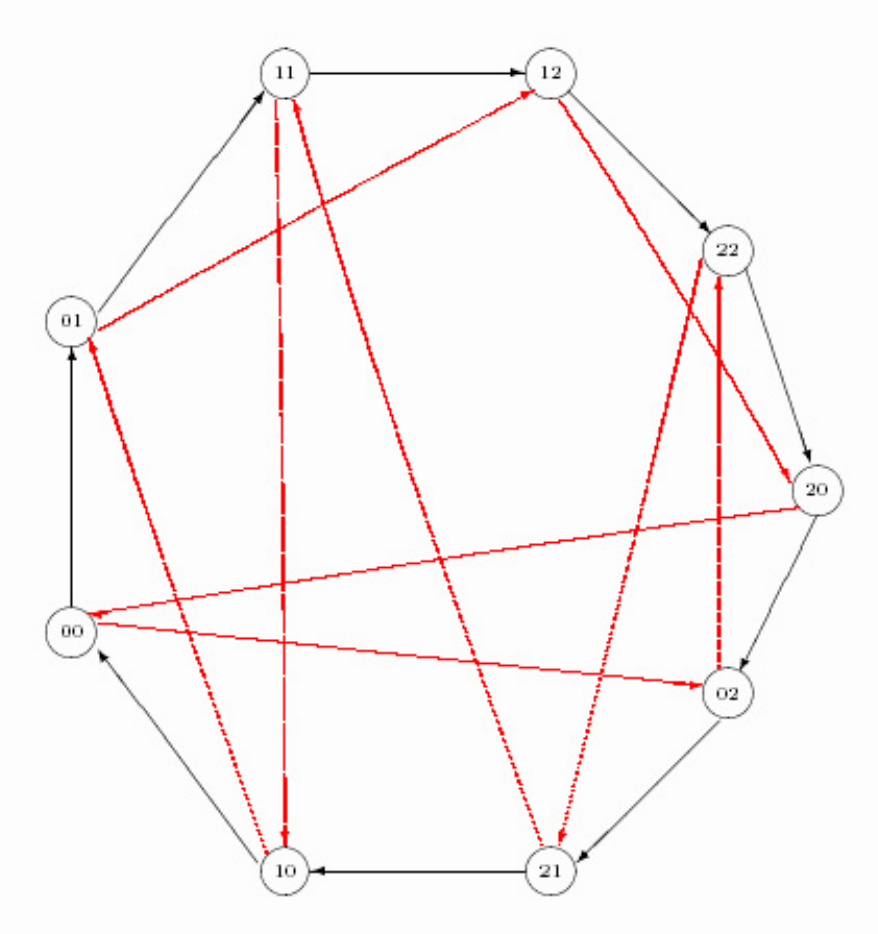}
\caption{Arc-disjoint Hamiltoniain cycles in De Bruijn graph $B(3,2)$}
\end{figure}

\newpage\section*{References}
          
\quad\ 1. M. C. Anisiu, Z. Bl\'azsik. Z. K\'asa, \textit{Maximal complexity of finite words}, 
Pure Math. and Appl. Vol. \textbf{13} (2002) No. 1--2, pp. 39--48.

\smallskip
      2. J. Bond, A. Iv\'anyi,  \textit{Modelling of interconnection
networks using De Bruijn graphs}, Third Conference of Program Designer, 
Ed. A. Iv\'anyi, Budapest, 1987,  75--87. 
{\scriptsize \url{http://www.acta.sapientia.ro/PD/Third-conference-of-program-designers.pdf}}

\smallskip
      3. N. G. de Bruijn, \textit{A combinatorial problem}, 
Nederl. Akad. Wetensch. Proc. 49 (1946), 758--764.

\smallskip
      4. Sophie Gire, 
\emph{R\'eseaux d'interconnexion},  Option Ma\^{\i}trise Informatique, 1996--97,
{\scriptsize {http://fastnet.univ-brest.fr/$\tilde{\ }$gire/COURS/OPTION$\_$RESEAUX/node48.html}}

\smallskip
      5. A. Iv\'anyi, M. Horv\'ath, \textit{Perfect sequences}, ICAI'2004, Eger, January 28, 2004.

\smallskip
      6. M. Lothaire, \textit{Combinatorics on words}, 
Addison-Wesley, Reading, MA, 1983.

\smallskip
      7. M. H. Martin, \textit{A problem in arrangements}, 
Bull. A.M.S. 40 (1934), 859--864.

\end{document}